\documentclass[journal,twoside,web]{ieeecolor}

\usepackage{generic}
\usepackage{cite}
\usepackage{amsmath,amssymb,amsfonts}
\usepackage{algorithmic}
\usepackage{graphicx}
\usepackage{textcomp}

\usepackage{graphicx}
\usepackage{subfigure}
\usepackage{epsfig} 
\usepackage{cite}
\usepackage{lcsys}

\usepackage{cite}
\usepackage{amsmath,amssymb,amsfonts,mathrsfs}
\usepackage{algorithmic}
\usepackage{graphicx}
\usepackage{textcomp}
\usepackage{mathtools}
\usepackage{commath}

\allowdisplaybreaks

\DeclareMathOperator{\sgn}{sgn}

\DeclareMathOperator{\esup}{ess\, sup}



\providecommand*{\ped}[1]{%
\ensuremath{_\textnormal{#1}}}

\providecommand*{\eu}%
{\ensuremath{\mathrm{e}}}
\providecommand*{\im}%
{\ensuremath{\mathrm{j}}}
\providecommand*{\GammaF}%
{\ensuremath{\mathrm{\Gamma}}}
\providecommand*{\BetaF}%
{\ensuremath{\mathrm{\Beta}}}

\newtheorem{theorem}{Theorem}[section]
\newtheorem{lemma}[theorem]{Lemma}

\begin{document}

\def\BibTeX{{\rm B\kern-.05em{\sc i\kern-.025em b}\kern-.08em
    T\kern-.1667em\lower.7ex\hbox{E}\kern-.125emX}}
\markboth{\journalname, VOL. XX, NO. XX, XXXX 2017}
{Romano \MakeLowercase{\textit{et al.}}: Linear viscoelastic rheological FrBD models}

\title{Linear viscoelastic rheological FrBD models}

\author{
Luigi Romano$^{a, b}$,
Ole Morten Aamo$^{b}$,
Jan Åslund$^{a}$,
Erik Frisk$^{a}$
\thanks{$^{a}$ Department of Electrical Engineering, Linköping University, Linköping, Sweden; \texttt{luigi.romano@liu.se}, \texttt{jan.aslund@liu.se}, \texttt{erik.frisk@liu.se}}
\thanks{$^{b}$ Department of Engineering Cybernetics, NTNU, Trondheim, Norway; \texttt{ole.morten.aamo@ntnu.no}}
}

\maketitle
\thispagestyle{empty}

\begin{abstract}
Recently, a new modeling paradigm, termed \emph{Friction with Bristle Dynamics} (FrBD), has been introduced for developing rate-and-state-dependent, control-oriented friction models. The framework combines nonlinear analytical expressions for the friction coefficient with constitutive equations for bristle-like elements. Within the FrBD framework, this letter introduces two novel formulations based on the two most general linear viscoelastic models for solids: the Generalized Maxwell (GM) and Generalized Kelvin-Voigt (GKV) elements. Both are analyzed in terms of boundedness and passivity, revealing that these properties are satisfied for any physically meaningful parametrization. An application of passivity for control design is also illustrated, considering an example from robotics. The findings of this letter systematically integrate rate-and-state dynamic friction models with linear viscoelasticity.
\end{abstract}


\section{Introduction}
\label{sec:introduction}

Friction plays a central role in the dynamics and control of mechanical systems, particularly in high-precision applications such as robotics, actuators, and mechatronic devices \cite{Motors,Hydraulics,C,Compensation}. Its complex behavior -- characterized by pre-sliding displacement, hysteresis, velocity weakening and strengthening effects, and memory -- has motivated the development of increasingly sophisticated dynamic friction models. Amongst these, rate-and-state-dependent descriptions, formulated in terms of nonlinear \emph{ordinary differential equations} (ODEs), have proven essential for capturing experimentally observed phenomena that cannot be reproduced by static or purely velocity-dependent laws \cite{Astrom1,Olsson,Bender,Integrated,Leuven}.

A prominent example is LuGre \cite{Astrom1,Olsson}, originally introduced in the context of control-oriented modeling to account for pre-sliding displacement and frictional lag through an internal bristle state. Since its introduction, the LuGre model and its variants have been widely adopted and extended in both theoretical and applied studies, particularly in control design and compensation schemes \cite{Motors,Hydraulics,C,Compensation,CanudasKelly}. Despite its success, the LuGre model exhibits well-known limitations, most notably the lack of passivity for general parameter choices and operating conditions, unless velocity-dependent damping terms are introduced \cite{Astrom2,Necessity,Hysteresis}. Additionally, the LuGre model involves several parameters whose physical interpretation is not straightforward, which complicates its use in multibody simulations \cite{Brogliato1}. This has prompted ongoing research into alternative formulations that preserve physical consistency whilst remaining amenable to analysis and control synthesis \cite{Bender,Integrated,Leuven}. 

Recently, a new modeling paradigm termed \emph{Friction with Bristle Dynamics} (FrBD) was introduced in \cite{FrBD} to systematically construct rate-and-state-dependent friction models by combining nonlinear friction laws with constitutive equations describing the internal bristle dynamics. This framework provides a unified and physically grounded approach that naturally incorporates rheological concepts from elasticity and viscoelasticity, allowing friction models to be interpreted as mechanical elements composed of springs and dashpots. Importantly, the FrBD framework enables the derivation of friction models that are inherently bounded and passive under physically meaningful parameterizations \cite{FrBD}, properties that are essential for stability and robustness in feedback control systems \cite{Passivity2,Passivity3}.

Within this context, the present letter extends the FrBD framework by introducing and analyzing a class of nonlinear viscoelastic models inspired by classical linear rheological elements. In particular, formulations based on the Generalized Maxwell (GM) and Generalized Kelvin-Voigt (GKV) representations are developed, allowing for richer descriptions of friction dynamics in solids. The mathematical properties of the resulting model are rigorously investigated, with particular emphasis on existence and uniqueness of solutions, boundedness, and passivity.
Beyond theoretical analysis, the letter also examines the qualitative dynamical behavior of the proposed models, showing that they reproduce key friction phenomena such as frictional hysteresis and relaxation behaviors observed in experimental studies. Finally, the practical relevance of the approach is demonstrated through a control application involving a robotic manipulator, where passivity of the friction model is sapiently exploited for output-feedback control design. This letter develops a unified framework combining general linear viscoelastic theory with rate-and-state-dependent friction laws, resulting in a family of dynamic models that admit clear physical interpretation and systematic characterization using standard engineering methods \cite{Sakh}. The proposed formulations are particularly relevant for applications involving highly viscoelastic materials, including polymers and rubbers. 


\subsection*{Notation and preliminaries}
In this letter, $\mathbb{R}$ denotes the set of real numbers; $\mathbb{R}_{>0}$ and $\mathbb{R}_{\geq 0}$ indicate the set of positive real numbers excluding and including zero, respectively. The set of positive integer numbers is indicated with $\mathbb{N}$, whereas $\mathbb{N}_{0}$ denotes the extended set of positive integers including zero, i.e., $\mathbb{N}_{0} = \mathbb{N} \cup \{0\}$.
$L^p((0,T);\mathbb{R}^n)$ and $C^k([0,T];\mathbb{R}^n)$ ($p, k \in \{1, 2, \dots, \infty\}$) denotes the spaces of $L^p$-integrable functions and $k$-times continuously differentiable functions on $[0,T]$ with values in $\mathbb{R}^n$ (for $T = \infty$, the closure of $[0,T]$ is identified with $\mathbb{R}_{\geq 0}$). For a function $f :(0,T) \mapsto \mathbb{R}$, the sup norm is defined as $\norm{f(\cdot)}_\infty \triangleq \esup_{(0,T)} \abs{f(\cdot)}$; $f : (0,T)\mapsto \mathbb{R}$ belongs to the space $L^\infty((0,T);\mathbb{R})$ if $\norm{f(\cdot)}_\infty < \infty$. A function $f \in C^0(\mathbb{R}_{\geq 0}; \mathbb{R}_{\geq 0})$ belongs to the space $\mathcal{K}$ if it is strictly increasing and $f(0) = 0$; $f \in \mathcal{K}$ belongs to the space $\mathcal{K}_\infty$ if it is unbounded. A function $f \in C^0(\mathbb{R}_{\geq 0}^2; \mathbb{R}_{\geq 0})$ belongs to the space $\mathcal{KL}$ if $f(\cdot,t) \in \mathcal{K}$ and is strictly decreasing in its second argument, with $\lim_{t\to \infty}f(\cdot,t) = 0$. 

Finally, an application of Theorem \ref{thm:Theorem1} below will guide the derivation of the friction models developed in the letter.

\begin{theorem}[Edwards \cite{Edwards}]\label{thm:Theorem1}
Suppose that the mapping $H : \mathbb{R}^{m+n}\mapsto \mathbb{R}^n$ is $C^1$ in a neighbourhood of a point $(x^\star,y^\star)$, where $H(x^\star,y^\star) = 0$. If the Jacobian matrix $\nabla_{y}H(x^\star,y^\star)^{\mathrm{T}}$ is nonsingular, there exist a neighbourhood $\mathcal{X}$ of $x^\star$ in $\mathbb{R}^m$, a neighbourhood $\mathcal{Y}$ of $(x^\star,y^\star)$ in $\mathbb{R}^{m+n}$, and a mapping $h \in C^1(\mathcal{X};\mathbb{R}^n)$ such that $y =h(x)$ solves the equation $H(y,x) = 0$ in $\mathcal{Y}$. 
In particular, the implicitly defined mapping $h(\cdot)$ is the limit of the sequence $\{h_k\}_{ k \in \mathbb{N}_0}$ of the successive approximations inductively defined by
\begin{subequations}
\begin{align}
h_{k+1}(x) & = h_k(x) - \nabla_{y}H(x^\star,y^\star)^{-\mathrm{T}}H\bigl(x,h_k(x)\bigr), \\
h_0(x) & = y^\star,
\end{align}
\end{subequations}
for $x\in \mathcal{X}$.
\end{theorem}

\section{Model derivation}\label{sect:Problem}
\begin{figure}
\centering
\includegraphics[width=0.55\linewidth]{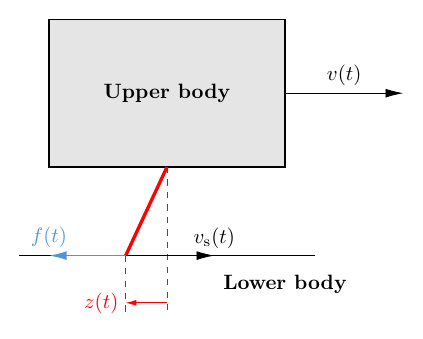} 
\caption{A schematic representation of the friction model}
\label{fig:LumpModel0}
\end{figure}

\begin{figure}
\centering
\includegraphics[width=0.9\linewidth]{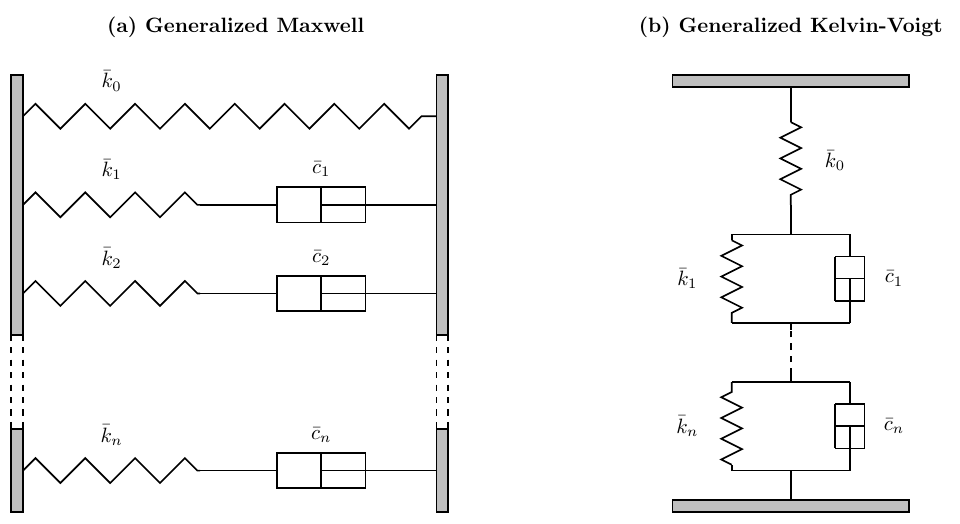} 
\caption{Linear viscoelastic rheological models for solid elements: \textbf{(a)} Generalized Maxwell (GM); \textbf{(b)} Generalized Kelvin-Voigt (GKV).}
\label{fig:Dashpot_GM}
\end{figure}

The present section recapitulates the FrBD framework introduced in \cite{FrBD}, which is here generalized to include more complete rheological descriptions of viscoelastic solids. 
To illustrate the salient aspects of the proposed approach, the situation illustrated schematically in Figure \ref{fig:LumpModel0} is considered: a rigid body travels with relative velocity $v \in \mathbb{R}$ with respect to a rigid substrate. To the lower boundary of the upper body, deformable bristles are attached, whose deflection is denoted by $z \in \mathbb{R}$. The total sliding velocity between the tip of the bristle and the lower body reads
\begin{align}
v\ped{s}(\dot{z},v) = v- \dot{z},
\end{align}
where Newton's notation has been adopted for the time derivative, i.e., $\dot{z} = \od{z}{t}$.
By deflecting, the bristle generates a force per unit of vertical load, $f \in \mathbb{R}$, which opposes the sliding motion. The dynamic relationship between the bristle deflection and the produced force must be appropriately specified by assuming a constitutive equation. Within the framework of linear viscoelasticity, the GM and GKV elements illustrated in Figure \ref{fig:Dashpot_GM} provide the most complete rheological descriptions of solids.
In particular, adopting a GM model with $n+1$ branches yields
\begin{subequations}\label{eq:fGM}
\begin{align}
f & = \bar{k}_0z + \sum_{i=1}^n f_i, \label{eq:GM_f0}\\
\dot{f}_i & = -\tau_i^{-1}f_i + \bar{k}_i\dot{z},
\end{align}
\end{subequations}
where $f_i \in \mathbb{R}$, $i \in \{1,\dots,n\}$, denote the internal forces generated by the dissipative branches, $\bar{k}_i \in \mathbb{R}_{>0}$, $i \in \{0,\dots,n\}$, are normalized micro-stiffness coefficients, and $\tau_i \in \mathbb{R}_{>0}$, $i \in \{1,\dots,n\}$, are relaxation time constants. On the other hand, using a GKV provides
\begin{subequations}\label{eq:GKV}
\begin{align}
f & = \bar{k}_0z -\bar{k}_0\sum_{i=1}^n z_i, \label{eq:GKV_f0}\\
f & = \bar{k}_iz_i + \bar{c}_i\dot{z}_i, 
\end{align}
\end{subequations}
where $z_i \in \mathbb{R}$, $i \in \{1,\dots,n\}$, represent the internal deformations of the dissipative branches, and $\bar{c}_i \in \mathbb{R}_{>0}$ are normalized micro-damping coefficients.

Equations \eqref{eq:fGM} and \eqref{eq:GKV} are equivalent after a suitable reparametrization, and admit a general representation in the form of the differential constitutive relationship \cite{Rheology2}
\begin{align}\label{eq:constDiff}
f + \sum_{i=1}^n \gamma_i \dod[i]{f}{t} = \sum_{i=0}^n \sigma_i \dod[i]{z}{t},
\end{align}
where the coefficients $\gamma_i$ and $\sigma_i$ can be determined from the original physical parameters. For what follows, \eqref{eq:constDiff} is reinterpreted algebraically as
\begin{align}\label{eq:constDiff2}
f\Bigl( \od[n]{z}{t}, \dots, \od{z}{t}, z, \od[n]{f}{t}, \dots, \od{f}{t}\Bigr) = \sum_{i=0}^n \sigma_i \dod[i]{z}{t}-\sum_{i=1}^n \gamma_i \dod[i]{f}{t}.
\end{align}

In the absence of inertial effects, the force $f$ in \eqref{eq:constDiff2} must counteract the nondimensional friction force exerted on the tip of the bristle:
\begin{align}\label{eq:ffric}
f\ped{r}\bigl(v\ped{s}(\dot{z},v)\bigr) = \dfrac{v\ped{s}(\dot{z},v)}{\abs{v\ped{s}(\dot{z},v)}_\varepsilon}\mu\bigl(v\ped{s}(\dot{z},v)\bigr),
\end{align}
where $\mu : \mathbb{R} \mapsto [\mu\ped{min},\infty)$, with $\mu\ped{min}\in \mathbb{R}_{>0}$, denotes the friction coefficient, and the function $\abs{\cdot}_\varepsilon \in C^0(\mathbb{R};\mathbb{R}_{\geq 0})$, with $\varepsilon \in \mathbb{R}_{\geq0}$, is a regularization of the absolute value $\abs{\cdot}$ for $\varepsilon\in \mathbb{R}_{>0}$, often converging uniformly to $\abs{\cdot}$ in $C^0(\mathbb{R};\mathbb{R}_{\geq 0})$ for $\varepsilon \to 0$ (e.g., $\abs{v}_\varepsilon= \sqrt{v^2 +\varepsilon}$), and with $\abs{\cdot}_\varepsilon \in C^1(\mathbb{R};\mathbb{R}_{\geq 0})$ for $\varepsilon \in \mathbb{R}_{>0}$ \cite{FrBD}. This letter assumes for simplicity $\mu\in C^0(\mathbb{R};[\mu\ped{min},\infty))$.

Equating \eqref{eq:constDiff2} with \eqref{eq:ffric} yields
\begin{align}\label{eq:nonlinImplGen}
\begin{split}
& H\Bigl( \od[n]{z}{t}, \dots, \od{z}{t}, z, \od[n]{f}{t}, \dots, \od{f}{t}, v\Bigr) \\
& \quad = f\Bigl( \od[n]{z}{t}, \dots, \od{z}{t}, z, \od[n]{f}{t}, \dots, \od{f}{t}\Bigr) - f\ped{r}\bigl(v\ped{s}(\dot{z},v)\bigr) = 0, \\
& \qquad \qquad \qquad \qquad \qquad \qquad \qquad \qquad \quad t \in (0,T),
\end{split}
\end{align}
and invoking Theorem \ref{thm:Theorem1} with $x \triangleq (z, \od[2]{z}{t}, \dots,\od[n]{z}{t}, \od{f}{t}, \dots, \od[n]{f}{t},v)$ and $y\triangleq \dot{z} = \od{z}{t}$ gives
\begin{align}\label{eq:zDotderGen}
\begin{split}
\dot{z}_{k+1} &= \dot{z}_k - {\dpd{H\Bigl( \od[n]{z}{t}^\star, \dots, \od{z}{t}^\star, z^\star, \od[n]{f}{t}^\star, \dots, \od{f}{t}^\star, v^\star\Bigr)}{\dot{z}}}^{-1} \\
& \quad \times H\Bigl( \od[n]{z}{t}, \dots, \od{z}{t}, z, \od[n]{f}{t}, \dots, \od{f}{t}, v\Bigr), \quad k \in \mathbb{N}_0. 
\end{split}
\end{align}
In turn, approximating \cite{FrBD,Rill}
\begin{align}\label{eq:approxDfGen}
\begin{split}
\dpd{H\Bigl( \od[n]{z}{t}, \dots, \od{z}{t}, z, \od[n]{f}{t}, \dots, \od{f}{t}, v\Bigr)}{\dot{z}} \approx  \sigma_1+ \dfrac{\mu\bigl(v\ped{s}(\dot{z},v)\bigr)}{\abs{v\ped{s}(\dot{z},v)}_\varepsilon},
\end{split}
\end{align}
recalling \eqref{eq:constDiff2}, and truncating \eqref{eq:zDotderGen} at $k=1$ gives, for an initial guess $z_0 =0$,
\begin{align}\label{eq:zValid}
\dot{z}(t) & = -\dfrac{\abs{v(t)}_\varepsilon}{\mu\bigl(v(t)\bigr)}f(t) + v(t), \quad t \in (0,T).
\end{align}
Within the linear viscoelastic framework, \eqref{eq:zValid} describes the evolution of the bristle dynamics independently of the assumed constitutive relationship. However, to completely characterize \eqref{eq:zValid}, it is again necessary to adopt a specific representation of the bristle element. Using \eqref{eq:fGM} yields the \textbf{FrBD}$_{n+1}$\textbf{-GM} model
\begin{subequations}\label{eq:fGM_ODE}
\begin{align}
\dot{z}(t) & = -\dfrac{\abs{v(t)}_\varepsilon}{\mu\bigl(v(t)\bigr)}\Biggl(\bar{k}_0z(t) + \sum_{i=1}^n f_i(t)\Biggr) + v(t) \\
\dot{f}_i(t) & = -\tau_i^{-1}f_i(t) + \bar{k}_i\dot{z}(t), \quad t \in (0,T), \; i \in \{1,\dots,n\}, \\
z(0) & = z_0, \; f_{i}(0)=f_{i,0}, \quad i \in \{1,\dots,n\},
\end{align}
\end{subequations}
whereas employing \eqref{eq:GKV} delivers the alternative \textbf{FrBD}$_{n+1}$\textbf{-GKV} formulation
\begin{subequations}\label{eq:GKV_ODE}
\begin{align}
\dot{z}(t) & = -\dfrac{\abs{v(t)}_\varepsilon}{\mu\bigl(v(t)\bigr)}\bar{k}_0\Biggl(z(t) - \sum_{i=1}^n z_i(t)\Biggr) + v(t) \\
\begin{split}
\dot{z}_i(t) & = -\dfrac{\bar{k}_i}{\bar{c}_i}z_i(t) + \dfrac{\bar{k}_0}{\bar{c}_i}\Biggl(z(t) - \sum_{i=1}^n z_i(t)\Biggr), \\
& \qquad \qquad \qquad \qquad \quad t \in (0,T), \; i \in \{1,\dots,n\}, 
\end{split} \\
z(0) & = z_0, \; z_{i}(0)=z_{i,0}, \quad i \in \{1,\dots,n\},
\end{align}
\end{subequations}
where the subscript $n+1$ stands for the dynamical order of the system (which coincides with the number of branches in the GM and GKV elements). Equations \eqref{eq:fGM_ODE} and \eqref{eq:GKV_ODE} generalize the Dahl, LuGre, and FrBD$_1$-KV model introduced in \cite{FrBD}.


\section{Model properties}\label{sect:Standard}
In this section, the FrBD$_{n+1}$ models \eqref{eq:fGM_ODE} and \eqref{eq:GKV_ODE} are analyzed in terms of well-posedness, boundedness, and passivity.
\subsection{Well-posedness and steady-state solution}
Theorem \ref{thm:wellPosedFrBD2SLS} asserts the well-posedness of \eqref{eq:fGM_ODE} and \eqref{eq:GKV_ODE} under standard assumptions. In the following, the state vector is defined as $\mathbb{R}^{n+1} \ni x(t) \triangleq [z(t)\; f_1(t) \; f_2(t) \; \dots f_n(t)]^{\mathrm{T}}$ and $\mathbb{R}^{n+1} \ni x(t) \triangleq [z(t)\; z_1(t) \; z_2(t) \; \dots z_n(t)]^{\mathrm{T}}$ for the FrBD$_{n+1}$-GM and FrBD$_{n+1}$-GKV variants, respectively.

\begin{theorem}[Existence and uniqueness of solutions]\label{thm:wellPosedFrBD2SLS}
For all inputs $v \in C^0([0,T];\mathbb{R})$ and ICs $x(0) = x_0 \in \mathbb{R}^{n+1}$, the FrBD$_{n+1}$ models \eqref{eq:fGM_ODE} and \eqref{eq:GKV_ODE} admit a unique solution $x \in C^1([0,T];\mathbb{R}^{n+1})$.
\end{theorem}
The results of Theorem \ref{thm:wellPosedFrBD2SLS} are of a qualitative nature. Formally, the transient solution to \eqref{eq:fGM_ODE} and \eqref{eq:GKV_ODE} may be recovered using the corresponding state-transition matrix after restating them in state-space representation. However, for large $n$, closed-form results are not obtainable. 
Instead, it is interesting to investigate the steady-state solution of \eqref{eq:fGM_ODE} and \eqref{eq:GKV_ODE}. Concerning the force, both variants yield
\begin{align}\label{eq:SSsol}
f = \sgn_\varepsilon(v)\mu(v),
\end{align}
with $\sgn_\varepsilon(v) \triangleq \frac{v}{\abs{v}_\varepsilon}$. The expressions for the bristle deformation and internal states instead differ for the FrBD$_{n+1}$-GM and FrBD$_{n+1}$-GKV models. Starting with \eqref{eq:fGM_ODE}, 
\begin{align}
z = \dfrac{f}{\bar{k}_0}, \quad f_i = 0, \quad i \in \{1,\dots,n\},
\end{align}
which coincides with the solution of the LuGre and FrBD$_1$-KV model introduced in \cite{FrBD}, for which $\sigma_0 = \bar{k}_0$. On the other hand, from \eqref{eq:GKV_ODE}, the following relationships are deduced:
\begin{align}\label{eq;ssGKM}
z = \sum_{i=0}^n \dfrac{f}{\bar{k}_i}, \quad z_i = \dfrac{f}{\bar{k}_i}, \quad i \in \{1,\dots,n\},
\end{align}
which are consistent with the parallel structure of the GKV.
In essence, \eqref{eq:SSsol}-\eqref{eq;ssGKM} demonstrate that the FrBD$_{n+1}$ models can reproduce any arbitrary steady-state friction characteristics in solids. 
The next Section \ref{sect:Math} discusses their dynamical features.

\subsection{Mathematical properties}\label{sect:Math}
This section investigates the boundedness and passivity properties of the FrBD$_{n+1}$ models \eqref{eq:fGM_ODE} and \eqref{eq:GKV_ODE}. Formal mathematical definitions of these notions are well established in the literature (see, e.g., \cite{Khalil}) and are therefore not repeated here. Instead, their physical interpretation in the context of friction modeling is briefly discussed. In particular, boundedness reflects the physically reasonable expectation that the friction force and internal states remain finite, coherently with \eqref{eq:ffric}. On the other hand, passivity captures the intrinsically dissipative nature of friction, and should be satisfied by any \emph{bona fide} friction model \cite{Necessity}. Essentially, it states that the power exchanged by the friction force must be nonnegative, that is, $pf(t)v(t) \geq 0$, where $p \in \mathbb{R}_{>0}$ is the normal component of the contact force. For the FrBD$_{n+1}$ models presented in this paper, both boundedness and passivity are always guaranteed.

Starting with the FrBD$_{n+1}$-GM variant, the following Lyapunov function candidate is considered:
\begin{align}\label{eq:LyapF}
V\bigl(x(t)\bigr) = \dfrac{1}{2}p\bar{k}_0z^2(t) + \dfrac{1}{2}\sum_{i=1}^n\dfrac{p}{\bar{k}_i}f_i^2(t).
\end{align}
Differentiating \eqref{eq:LyapF} along the dynamics \eqref{eq:fGM_ODE}, and using \eqref{eq:GM_f0}, yields
\begin{align}\label{eq:LyapFInal}
\begin{split}
\dot{V}(t) & = pf(t)v(t)-\dfrac{p\abs{v(t)}_\varepsilon}{\mu\bigl(v(t)\bigr)}f^2(t) \\
& \quad -\sum_{i=1}^n \dfrac{p}{\tau_i \bar{k}_i}f_i^2(t), \quad t \in (0,T).
\end{split}
\end{align}
For the FrBD$_{n+1}$-GKV model, it is convenient to introduce the variable $\mathbb{R}\ni z_0(t) \triangleq z(t) - \sum_{i=1}^n z_i(t)$, whose dynamics obeys
\begin{subequations}
\begin{align}
\dot{z}(t) & = -\dfrac{\abs{v(t)}_\varepsilon}{\mu\bigl(v(t)\bigr)}\bar{k}_0z_0(t)-\sum_{i=1}^n \dot{z}_i(t) + v(t), \quad t \in(0,T), \\
z_0(0) & = z_{0,0}.
\end{align}
\end{subequations}
Accordingly, the following Lyapunov function candidate is considered:
\begin{align}\label{eq:LyapF2}
V\bigl(x(t)\bigr) = \dfrac{1}{2}p\bar{k}_0z_0^2(t) + \dfrac{1}{2}\sum_{i=1}^n p\bar{k}_i z_i^2(t).
\end{align}
Differentiating \eqref{eq:LyapF2} along the dynamics \eqref{eq:GKV_ODE}, and using \eqref{eq:GKV_f0}, gives
\begin{align}\label{eq:LyapFInal2}
\begin{split}
\dot{V}(t) & = pf(t)v(t)-\dfrac{p\abs{v(t)}_\varepsilon}{\mu\bigl(v(t)\bigr)}f^2(t) \\
& \quad -\sum_{i=1}^n \dfrac{p}{\bar{c}_i}\bigl(\bar{k}_iz_i(t)-\bar{k}_0z_0(t)\bigr)^2, \quad t \in (0,T).
\end{split}
\end{align}
The above \eqref{eq:LyapFInal} and \eqref{eq:LyapFInal2} will be instrumental to prove boundedness and passivity for the FrBD$_{n+1}$ models \eqref{eq:fGM_ODE} and \eqref{eq:GKV_ODE}. The main results in this sense are enounced in the following Sections \ref{sect:Bound} and \ref{sect:pass}.
\subsubsection{Boundedness}\label{sect:Bound}
The next result, formalized in Lemma \ref{lemma:boundFrBD-SLS}, is concerned with the boundedness of the FrBD$_{n+1}$ models.
\begin{lemma}[Boundedness]\label{lemma:boundFrBD-SLS}
Suppose that $\abs{y}_\varepsilon \geq \abs{y}$ for every $y \in \mathbb{R}$. Then, for all inputs $v \in C^0(\mathbb{R}_{\geq 0};\mathbb{R})\cap L^\infty(\mathbb{R}_{\geq 0};\mathbb{R})$ and ICs $x_0 \in \mathbb{R}^{n+1}$, the solutions of \eqref{eq:fGM_ODE} and \eqref{eq:GKV_ODE} are bounded for all times. 

\begin{proof}
For brevity, the proof is detailed considering the FrBD$_{n+1}$-GM formulation; the corresponding result for the FrBD$_{n+1}$-GKV variant follows from identical arguments.

From \eqref{eq:LyapF}, it follows the existence of $\gamma_1 \in \mathbb{R}_{>0}$ such that
\begin{align}\label{eq:LyapFInalP1}
\begin{split}
\dot{V}(t) & = \gamma_1 \abs{v(t)}_\varepsilon\sqrt{V(t)}-\dfrac{p\abs{v(t)}_\varepsilon}{\mu\bigl(v(t)\bigr)}f^2(t)  \\
& \quad -\sum_{i=1}^n \dfrac{p}{\tau_i \bar{k}_i}f_i^2(t), \quad t \in (0,T).
\end{split}
\end{align}
Moreover, by defining $\mathbb{R}_{>0}\ni \varrho_1  \triangleq \max_{t\in \mathbb{R}_{\geq 0}} \abs{v(t)}_\varepsilon$ and $\mathbb{R}_{>0}\ni \varrho_2  \triangleq \max_{t\in \mathbb{R}_{\geq 0}}\mu\bigl(v(t)\bigr)$, it may be inferred that
\begin{align}\label{eq:LyapFInalP2}
\begin{split}
\dot{V}(t) & = \gamma_1 \abs{v(t)}_\varepsilon\sqrt{V(t)} \\
& \quad -p\abs{v(t)}_\varepsilon\Biggl(\dfrac{f^2(t) }{\varrho_2} +\sum_{i=1}^n\dfrac{1}{\varrho_1\tau_i\bar{k}_i}f_i^2(t)\Biggr) \\
& \leq -\abs{v(t)}_\varepsilon\sqrt{V(t)}\Bigl(\gamma_2\sqrt{V(t)}-\gamma_1\Bigr), \quad t \in (0,T).
\end{split}
\end{align}
for some $\gamma_2 \in \mathbb{R}_{>0}$. For all $\sqrt{V}(t) \geq \frac{\gamma_1}{\gamma_2}$, \eqref{eq:LyapFInalP2} implies $V(t) \leq \max\{V(0), \frac{\gamma_1}{\gamma_2}\}$ for all $t \in [0,T]$, from which the claim follows.
\end{proof}
\end{lemma}

\subsubsection{Passivity}\label{sect:pass}
Passivity is asserted by Lemma \ref{lemma:PassFrBDSls} below.
\begin{lemma}[Passivity]\label{lemma:PassFrBDSls}
Consider the FrBD$_{n+1}$ models \eqref{eq:fGM_ODE} and \eqref{eq:GKV_ODE} with input $v \in C^0([0,T];\mathbb{R})$ and output $pf \in C^0([0,T];\mathbb{R})$. Then, the mapping $\Sigma : v \mapsto pf$ is passive with storage functions \eqref{eq:LyapF} and \eqref{eq:LyapF2}, respectively.

\begin{proof}
Rearranging \eqref{eq:LyapFInal} and \eqref{eq:LyapFInal2} and integrating over time immediately yields
\begin{align}
\int_0^t pf\bigl(t^\prime\bigr)v\bigl(t^\prime\bigr) \dif t^\prime \geq V\bigl(x(t))-V(x_0), \quad t \in [0,T],
\end{align}
with $V(x(t))$ reading respectively as in \eqref{eq:LyapF} and \eqref{eq:LyapF2}.
\end{proof}
\end{lemma}

\section{Dynamical behavior and control applications}\label{sect:appl}
The dynamics of the FrBD$_{n+1}$ models is investigated qualitatively in Section \ref{sect:dyn}, whereas a control application related to a robotic manipulator is then exemplified in Section \ref{sect:control}.
\subsection{Dynamical behavior}\label{sect:dyn}
The dynamical behavior of the FrBD models is discussed concerning \emph{frictional lag} and \emph{relaxation}.

\subsubsection{Frictional lag}

Hess and Soom \cite{Hess} investigated the velocity-dependent dynamic behavior of friction during unidirectional motion. Their experiments demonstrated the presence of hysteresis in the friction force-velocity relationship: for a given velocity magnitude, the friction force is lower during deceleration than during acceleration. In addition, the width of the hysteresis loop was observed to vary with the rate of change of velocity, expanding or contracting as this rate increases. To assess whether the FrBD model reproduces this characteristic, a first set of simulations was performed by postulating the friction coefficient as
\begin{align}\label{eq:fViscOOO}
\mu(v\ped{s}) = \mu\ped{d} + (\mu\ped{s}-\mu\ped{d})\eu^{-(\abs{v\ped{s}}/v\ped{S})^{\delta\ped{S}}},
\end{align}
where $\mu\ped{d}, \mu\ped{s} \in \mathbb{R}_{>0}$ denote the \emph{dynamic} and \emph{static friction coefficient}, respectively, $v\ped{S}\in \mathbb{R}_{\geq 0}$ is the \emph{Stribeck velocity}, $\delta\ped{S} \in \mathbb{R}_{\geq 0}$ the \emph{Stribeck exponent}.

\begin{figure}
\centering
\includegraphics[width=0.85\linewidth]{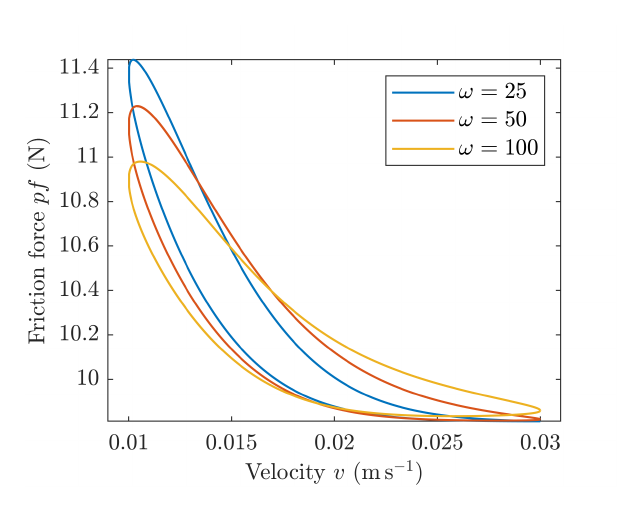} 
\caption{Frictional lag hysteresis.}
\label{fig:Hyst}
\end{figure}

For relatively high values of the damping coefficient $\sigma_1$, the hysteresis loop is generally inverted when the Stribeck velocity $v\ped{S}$ is sufficiently large. However, the conventional hysteretic behavior is recovered for sufficiently small values of $v\ped{S}$. Figure \ref{fig:Hyst} shows the hysteresis curves obtained by imposing relative velocities at different excitation frequencies, $\omega = 25$, 50, and 100 Hz. As the frequency increases, the peak friction force $pf$ decreases, whilst the hysteresis loops become progressively broader, consistent with the preceding discussion. Figure \ref{fig:Hyst} was generated using the model parameters listed in Table \ref{tab:param2}, with $\sigma_0 =\bar{k}_1$, $\sigma_1 = \tau_1(\bar{k}_0 + \bar{k}_1)$, and $\gamma_1 = \tau_1$ in \eqref{eq:constDiff} according to the GM parametrization.

\begin{table}[h!]\centering 
\caption{Model parameters}
{\begin{tabular}{|c|c|c|c|}
\hline
Parameter & Description & Unit & Value \\
\hline 
$\sigma_0$ & Normalized micro-stiffness & $\textnormal{m}^{-1}$ & $10^4$ \\
$\sigma_1$ & Normalized micro-damping & $\textnormal{s}\,\textnormal{m}^{-1}$ &64.5 \\
$\gamma_1$ & Relaxation time & s &0.001 \\
$\mu\ped{d}$ & Dynamic friction coefficient & - &1 \\
$\mu\ped{s}$ & Static friction coefficient & - &1.5 \\
$v\ped{S}$ & Stribeck velocity & $\textnormal{m}\,\textnormal{s}^{-1}$ &0.01 \\
$\delta\ped{S}$ & Stribeck exponent & - &2 \\
$\varepsilon$ & Regularization parameter & $\textnormal{m}^2\,\textnormal{s}^{-2}$ & $0$ \\
\hline
\end{tabular} }
\label{tab:param2}
\end{table}

\subsubsection{Relaxation}
\begin{figure}
\centering
\includegraphics[width=0.85\linewidth]{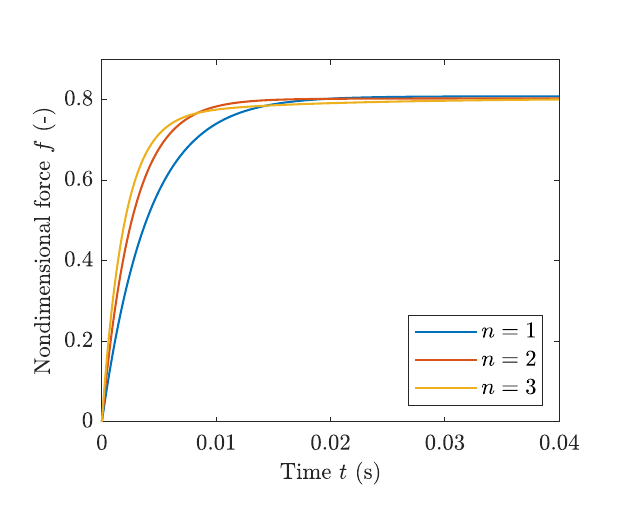} 
\caption{Relaxation behavior for different rheological orders.}
\label{fig:Rel}
\end{figure}

When a constant relative sliding velocity is imposed between a viscoelastic body and a rigid substrate, the generated friction force does not instantaneously attain its steady-state value. Instead, a transient phase is observed before the stationary friction level is fully established. This phenomenon is commonly referred to as \emph{relaxation behavior}. For highly viscoelastic materials, such as the polymers typically used in tyres, internal stress relaxation occurs over a broad frequency spectrum. Accurately capturing this behavior requires rheological models that incorporate multiple relaxation times \cite{Rheology2}.
Figure \ref{fig:Rel} illustrates the relaxation behavior predicted by the FrBD$_{n+1}$-GM models for $n=1$, 2, and 3, with parameters as in Table \ref{tab:param2}, $\tau_1 = 1$, $\tau_2 = 0.1$, $\tau_3 = 0.01$ s and $v =0.01$ $\textnormal{m}\,\textnormal{s}^{-1}$. According to \eqref{eq:SSsol} and \eqref{eq:fGM_ODE}, all the variants yield the same steady-state response, but their relaxation dynamics is influenced by the rheological order.

\subsection{Feedback control of a robotic arm}\label{sect:control}\color{black}
Inspired by \cite{CanudasKelly}, the following example elucidates an application of Lemma \ref{lemma:PassFrBDSls} for output-feedback control design. To this end, the following ODE system, describing a 1-DOF\footnote{The results of this section may be easily extended to the case of multi-DOF systems.} robot manipulator with FrBD$_{n+1}$-GM friction at the joint, is considered \cite{CanudasKelly}: 
\begin{subequations}\label{eq:ODEarm}
\begin{align}
\begin{split}
& m\bigl(q(t)\bigr)\ddot{q}(t) = - c\bigl(q(t),\dot{q}(t)\bigr)\dot{q}(t) - \mathrm{g}\bigl(q(t)\bigr) \\
& \qquad \qquad \qquad   - F(t) + U(t), \label{eq:fristEq}
\end{split}\\
& \dot{z}(t) = - \dfrac{\abs{\dot{q}(t)}_\varepsilon}{\mu\bigl(\dot{q}(t)\bigr)}\Biggl(\bar{k}_0z(t) + \sum_{i=1}^n f_i(t)\Biggr)+ \dot{q}(t), \label{eq:zDerSLSq}\\
& \dot{f}_i(t) = -\tau_i^{-1}f_i(t) + \bar{k}_i\dot{z}(t), \quad t \in (0,T),\; i \in \{1,\dots,n\},\label{eq:zDerSLSfq} \\
\begin{split}
& q(0) = q_0, \; \dot{q}(0) = \dot{q}_0, \\
& z(0) = z_0, \; f_{i}(0)=f_{i,0}, \quad i \in \{1,\dots,n\},
\end{split}
\end{align}
\end{subequations}
where $q(t) \in \mathbb{R}$ denotes the position of the robotic arm, $U(t) \in \mathbb{R}$ the control input, and $\mathbb{R} \ni F(t) \triangleq rpf(t)$ is the friction torque, being $r \in \mathbb{R}_{>0}$ a reference joint radius. In \eqref{eq:fristEq}, $m(q(t)) \in \mathbb{R}_{>0}$ is the mass term, $c(q(t),\dot{q}(t))\dot{q}(t) \in \mathbb{R}$ accounts for the Coriolis and centripetal effects, and $g(q(t)) \in \mathbb{R}$ is the gravitational torque.
Assuming that $q(t)$ and $\dot{q}(t)$ are measured, the objective consists of tracking a reference position $q\ped{ref} \in C^2(\mathbb{R}_{\geq 0}; \mathbb{R})$. Defining $\mathbb{R} \ni \tilde{q}(t) \triangleq q(t)-q\ped{ref}(t)$ and $\mathbb{R} \ni s(t) \triangleq \dot{\tilde{q}}(t) + \lambda \tilde{q}(t)$, the control input is specified as \cite{CanudasKelly}
\begin{align}\label{eq:Uinput}
\begin{split}
U(t) & = -k_1s(t) + \hat{F}(t)+m\bigl(q(t)\bigr)\bigl[\ddot{q}\ped{ref}(t)-\lambda\dot{\tilde{q}}(t)\bigr] \\
& \quad +c\bigl(q(t),\dot{q}(t)\bigl)\bigl[\dot{q}\ped{ref}(t)-\lambda\tilde{q}(t)\bigr] + \mathrm{g}\bigl(q(t)\bigr),
\end{split}
\end{align}
where $\lambda, k_1 \in \mathbb{R}_{>0}$ are control gains to be selected, and $\mathbb{R} \ni \hat{F}(t) \triangleq rp\hat{f}(t)$ is an estimate of $F(t)$ obtained using the observer
\begin{subequations}\label{eq:FrBD-SDSObs}
\begin{align}
& \dot{\hat{z}}(t) = - \dfrac{\abs{\dot{q}(t)}_\varepsilon}{\mu\bigl(\dot{q}(t)\bigr)}\Biggl(\bar{k}_0\hat{z}(t) + \sum_{i=1}^n \hat{f}_i(t)\Biggr)+ \dot{q}(t)-k_2s(t), \label{eq:zDerSLSObs}\\
& \dot{\hat{f}}_i(t) = -\tau_i^{-1}\hat{f}_i(t) + \bar{k}_i\dot{\hat{z}}(t), \quad t \in (0,T),\; i \in \{1,\dots,n\},\label{eq:zDerSLSfObs} \\
& \hat{z}(0) = \hat{z}_0, \; \hat{f}_i(0) = \hat{f}_{i,0}, \quad i \in \{1,\dots,n\},
\end{align}
\end{subequations}
where $k_2 \in \mathbb{R}_{>0}$ denotes the observer gain.

Consequently, introducing the error variables $\mathbb{R}\ni \tilde{z} \triangleq z-\hat{z}(t)$, $\mathbb{R}\ni \tilde{f}(t) \triangleq f(t)-\hat{f}(t)$, and $\mathbb{R}\ni \tilde{F}(t) \triangleq F(t)-\hat{F}(t) = rp\tilde{f}(t)$, and using \eqref{eq:Uinput}, the following system is obtained:
\begin{subequations}
\begin{align}
& m\bigl(q(t)\bigr)\dot{s}(t) = -c\bigl(q(t),\dot{q}(t)\bigr)s(t)-k_1s(t)-\tilde{F}(t), \label{4a}\\
& \dot{\tilde{z}}(t) = - \dfrac{\abs{\dot{q}(t)}_\varepsilon}{\mu\bigl(\dot{q}(t)\bigr)}\Biggl(\bar{k}_0\tilde{z}(t) + \sum_{i=1}^n \tilde{f}_i(t)\Biggr)+k_2s(t), \label{eq:zDerSLSObsErr}\\
& \dot{\tilde{f}}_i(t) = -\tau_i^{-1}\tilde{f}_i(t) + \bar{k}_i\dot{\tilde{z}}(t), \quad t \in (0,T),\; i \in \{1,\dots,n\},\label{eq:zDerSLSfObsErr} \\
& s(0) = s_0, \; \tilde{z}(0) = \tilde{z}_0, \; \tilde{f}_i(0) = \tilde{f}_{i,0}, \quad i \in \{1,\dots,n\}.
\end{align}
\end{subequations}
From Proposition 2 in \cite{CanudasKelly}, it follows that the operator $\mathrm{\Sigma}_1 : -\tilde{F}\mapsto s$ defined by \eqref{4a} is output-strictly-passive, that is, there exists $\beta_0\in \mathbb{R}_{\geq 0}$ and $\beta_1 \in \mathbb{R}_{>0}$ such that
\begin{align}\label{eq:sPass1}
- \int_0^t \tilde{F}\bigl(t^\prime\bigr)s\bigl(t^\prime\bigr) \dif t^\prime \geq -\beta_0+ \beta_1\int_0^t \abs{s\bigl(t^\prime\bigr)}^2\dif t^\prime, \quad t\in\mathbb{R}_{\geq 0}.
\end{align}
Moreover, by Lemma \ref{lemma:PassFrBDSls}, the operator $\mathrm{\Sigma}_2 : s \mapsto \tilde{F}$ defined by \eqref{eq:zDerSLSObsErr}, \eqref{eq:zDerSLSfObsErr}, and $\tilde{F} = rp\tilde{f}$ is passive. In particular, there exist $\beta_2 \in \mathbb{R}_{\geq 0}$ such that
\begin{align}\label{eq:sPass2}
\int_0^t \tilde{F}\bigl(t^\prime\bigr)s\bigl(t^\prime\bigr) \dif t^\prime \geq -\beta_2, \quad t\in\mathbb{R}_{\geq 0}.
\end{align}
Combining \eqref{eq:sPass1} and \eqref{eq:sPass2} implies $s \in L^2(\mathbb{R}_{>0};\mathbb{R})$. Invoking similar arguments as in \cite{CanudasKelly}, it may be finally concluded that $\lim_{t \to \infty} \tilde{q}(t) = 0$. The claim, formalized in Theorem \ref{themFeedback} below, is the logical consequence of known results on passive systems interconnections \cite{Brogliato2}. 
\begin{theorem}\label{themFeedback}
Consider the ODE system \eqref{eq:ODEarm} with the observer \eqref{eq:FrBD-SDSObs}. Then, the control input \eqref{eq:Uinput} ensures that $\lim_{t \to \infty} \tilde{q}(t) = 0$ for all $(q_0,\dot{q}_0,z_0,f_{1,0},\dots, f_{n,0}) \in \mathbb{R}^{n+3}$. 
\end{theorem}
An analogous result can be obtained by replacing the FrBD$_{n+1}$-GM friction model \eqref{eq:fGM_ODE} in \eqref{eq:ODEarm} with the its FrBD$_{n+1}$-GKV counterpart described by \eqref{eq:GKV_ODE}.

\section{Conclusion}\label{sect:concl}
This letter presented a class of linear viscoelastic friction models within the Friction with Bristle Dynamics (FrBD) framework, based on the two most general rheological descriptions of solids: the Generalized Maxwell (GM) and Generalized Kelvin-Voigt (GKV) elements. 

Utilizing a natural energy-based Lyapunov interpretation, the proposed models, formulated as systems of $n+1$ nonlinear ODEs, were shown to be well-posed and to satisfy boundedness and passivity for all parametrizations. Qualitative analysis demonstrated that key friction and rheological phenomena, including frictional hysteresis and stress relaxation, are accurately reproduced and consistent with experimental observations and established dynamic models. Finally, a robotic control example illustrated how the passivity of the FrBD model can be effectively exploited in output-feedback control design.
These findings confirm the FrBD framework as a physically grounded and control-oriented approach to dynamic friction modeling. In particular, the results of this letter unify, for the first time, linear viscoelasticity and rate-and-state friction models. Future research will address experimental validation and explore nonlinear rheological extensions and integration within multibody simulations.


\end{document}